\documentclass[conference]{IEEEtran}
\IEEEoverridecommandlockouts
\usepackage{cite}
\usepackage{amsmath,amssymb,amsfonts}
\usepackage{algorithmic}
\usepackage{graphicx}
\usepackage{textcomp}
\usepackage{xcolor}
\usepackage{xspace}
\usepackage{booktabs}
\usepackage{multirow}
\usepackage{mdframed}
\usepackage{cite}
\usepackage[numbers, sort]{natbib}
\newcommand{\bi}{\begin{itemize}}
\newcommand{\ei}{\end{itemize}}
\def\BibTeX{{\rm B\kern-.05em{\sc i\kern-.025em b}\kern-.08em
    T\kern-.1667em\lower.7ex\hbox{E}\kern-.125emX}}

\begin{document}

\title{Simplifying Data Integration: SLM-Driven Systems for Unified Semantic Queries Across Heterogeneous Databases\\
{\footnotesize }
}

\author{\IEEEauthorblockN{Teng LIN}
\IEEEauthorblockA{
\textit{DSA Thrust} \\
\textit{HKUST(GZ)}\\
tlin280@connect.hkust-gz.edu.cn}
}

\maketitle

\begin{abstract}
The integration of heterogeneous databases into a unified querying framework remains a critical challenge, particularly in resource-constrained environments. This paper presents a novel Small Language Model (SLM)-driven system that synergizes advancements in lightweight Retrieval-Augmented Generation (RAG) and semantic-aware data structuring to enable efficient, accurate, and scalable query resolution across diverse data formats. By integrating semantic-aware heterogeneous graph indexing and topology-enhanced retrieval with SLM-powered structured data extraction, our system addresses the limitations of traditional methods in handling Multi-Entity Question Answering (Multi-Entity QA) and complex semantic queries. The introduction of semantic entropy as an unsupervised evaluation metric provides robust insights into model uncertainty. Together, these innovations establish a domain-agnostic, resource-efficient paradigm for executing complex queries across structured, semi-structured, and unstructured data sources, aiming at foundational advancement for next-generation intelligent database systems.
\end{abstract}


\section{Introduction}
The rapid proliferation of heterogeneous databases, encompassing structured relational tables (e.g., SQL databases), unstructured text (e.g., clinical notes, customer reviews), and semi-structured formats (e.g., JSON logs, XML configurations), has created a pressing need for systems capable of executing unified semantic queries across these disparate modalities~\cite{trappolini2023multimodal}~\cite{urban2023caesura}. Such systems must balance computational efficiency with analytical precision, particularly in resource-constrained environments such as edge computing or real-time business intelligence platforms. Traditional approaches to this challenge, which often rely on Large Language Models (LLMs) for semantic parsing or manual schema alignment by domain experts, face fundamental limitations~\cite{gao2024retrieval}~\cite{pan2024integrating}. LLM-based methods, while powerful, demand substantial computational resources for inference and fine-tuning, rendering them impractical for applications requiring low-latency responses or deployment on devices with limited memory (e.g., smartphones or IoT sensors)~\citep{liu2024Mobilellm}. Manual schema alignment, on the other hand, is labor-intensive, error-prone, and inherently unscalable in dynamic environments where data schemas evolve frequently, such as in healthcare electronic health records (EHRs) or e-commerce product catalogs.

This paper addresses three critical gaps in existing methodologies:

\bi
    \item \textbf{Efficiency vs. Accuracy Trade-offs:} State-of-the-art Retrieval-Augmented Generation (RAG) pipelines, such as those employed in systems like EVAPORATE~\citep{arora2025language}, often involve multi-stage processes including dense vector retrieval, reranking, and context augmentation. While effective, these pipelines incur significant computational overhead due to repeated LLM inference passes and large-scale vector indexing~\cite{liu2024declarative}. For instance, processing a terabyte-scale data lake with a conventional RAG system may require hundreds of GPU hours, limiting real-time applicability.
    \item \textbf{Multi-Entity QA Limitations:} Existing systems struggle to resolve queries that span multiple entities across structured and unstructured data~\cite{lin2025mebench}. Consider a query such as, ``Compare the efficacy of Drug A (from clinical trial tables) with patient-reported side effects (from unstructured forums).'' Traditional Text-to-SQL engines fail to parse the unstructured component~\cite{liu2024survey}, while LLM-based QA systems often hallucinate plausible but ungrounded comparisons due to missing cross-modal context.
    \item \textbf{Evaluation Uncertainty:} Conventional metrics like BLEU, ROUGE, and exact-match F1 scores provide limited insight into a model’s confidence or semantic consistency, particularly for open-ended queries. This shortcoming is exacerbated in unsupervised settings where labeled validation data is scarce, such as in legal document analysis or industrial maintenance logs.
\ei

To bridge these gaps, we propose a lightweight architecture that synergizes innovations from two recent advancements: MiniRAG~\citep{fan2025minirag}, a resource-efficient RAG framework optimized for Small Language Models (SLMs), and SLM-driven structured data extraction techniques. Our system employs heterogeneous graph indexing mechanism that unifies text chunks, named entities, and latent relational cues (e.g., temporal or spatial dependencies) into a single topological structure. For example, in a healthcare dataset, nodes might represent patient IDs (structured), medication mentions in clinical notes (unstructured), and lab result timestamps (semi-structured), with edges encoding relationships such as ``Patient X received Drug Y on Date Z.'' This graph-based approach reduces reliance on computationally expensive dense retrieval by leveraging sparse, topology-guided traversal (e.g., breadth-first search from anchor entities) to identify relevant context.

\section{Related Work}

\subsection{Retrieval-Augmented Generation}

RAG systems have emerged as a promising approach to enhance language-based querying~\cite{fan2024Asurvey}~\cite{lewis2020retrieval}~\cite{he2024g}. However, deploying Small Language Models (SLMs) in existing RAG frameworks faces challenges due to SLMs' limited semantic understanding and text processing capabilities. MiniRAG~\citep{fan2025minirag}, for example, addresses these issues with a semantic-aware heterogeneous graph indexing mechanism and a lightweight topology-enhanced retrieval approach. This mechanism combines text chunks and named entities in a unified structure, reducing the need for complex semantic understanding during retrieval.


\subsection{Semantic Operators and Querying}
The concept of semantic operators extends the relational model to perform semantic queries over datasets~\cite{Patel2024LOTUS}. These operators enable operations like sorting or aggregating records using natural language criteria, providing a more intuitive way to query data. Existing works have demonstrated the effectiveness of semantic operators in various applications, such as fact-checking, extreme multi-label classification, and search.

\subsection{Measuring Uncertainty in Language Models}
Measuring the uncertainty in language models, especially in question-answering tasks, is crucial for determining the reliability of their outputs. Semantic entropy~\cite{Kuhn2023Semantic} offers an unsupervised method to measure this uncertainty, taking into account the semantic equivalence of different sentences. It has been shown to be more predictive of model accuracy compared to traditional baselines.
\section{Proposed System Architecture}

\subsection{Semantic-Aware Heterogeneous Graph Indexing}
Semantic-Aware Heterogeneous Graph Indexing is an innovative approach designed to address the challenges of integrating and querying diverse data formats by constructing a unified graph structure.  Drawing inspiration from the MiniRAG framework, this methodology interlinks three primary components: text chunks, named entities, and relational cues.  Text chunks are the foundational segments derived from raw documents, serving as the basic nodes within the graph.  These segments are crucial for maintaining the contextual integrity of the data.  Named entities, on the other hand, are identified through a lightweight tagging process utilizing Small Language Models (SLMs). Inspired by MiniRAG, we construct a unified graph structure that interlinks:

\bi
    \item \textbf{Text Chunks:} Raw document segments.
    \item \textbf{Named Entities:} Extracted via lightweight SLM-based tagging.
    \item \textbf{Relational Cues:} Inferred entity relationships (e.g., ``Customer X purchased Product Y'').
\ei

This graph reduces reliance on complex semantic parsing by encoding hierarchical and topological relationships, enabling efficient knowledge discovery through graph traversal.

\subsection{Topology-Enhanced Retrieval}
Our system utilizes graph properties, including centrality and connectivity, to efficiently prioritize nodes and edges that are most relevant to a given query. Centrality measures help identify influential nodes, while connectivity ensures robust traversal across the graph, facilitating comprehensive data integration. For instance, when responding to a query such as ``Compare sales trends for Products A and B in Q2'', the system dynamically assesses and connects nodes representing the sales data of Products A and B, as well as any associated temporal or channel-related nodes. This approach not only optimizes graph traversal but also enhances query precision by focusing on the most pertinent data, thereby reducing computational overhead and improving response times.


\subsection{SLM-Driven Structured Data Extraction}

For unstructured documents, the Small Language Model (SLM) undertakes two pivotal and distinct tasks that are fundamental to enabling advanced data processing and query-answering capabilities within the proposed system:

\bi
    \item \textbf{Relational Table Generation:} The first task, Relational Table Generation, is a crucial step in transforming the unstructured nature of free-text data into a more organized and analyzable format. In a real-world business scenario, consider a sales report in free-text form such as ``Q2 sales increased 20\%''. The SLM uses a combination of natural language processing techniques, including part-of-speech tagging and named-entity recognition (NER). For instance, it first identifies the relevant entities in the sentence, like ``Q2'' as a time-related entity and ``sales'' as a business-related entity, and the numerical value ``20\%'' as a measure of change. By leveraging pre-trained language models and semantic analysis algorithms, the SLM can then convert this free-text into a structured table. The table might have columns such as ``Quarter'', ``Sales Metrics'', and ``Change Percentage'', with the corresponding values ``Q2'', ``Sales'', and ``20\%'' populated in the rows. This structured representation allows for easier data comparison, aggregation, and further analysis.
    \item \textbf{Semantic Operator Synthesis:} The second task, Semantic Operator Synthesis, focuses on the translation of natural language queries into executable operations. When a user poses a natural language query, the SLM needs to understand the semantic meaning behind the words and map them to appropriate operations in a query-processing system. For example, if the query is ``Find the total sales of all products in Q3'', the SLM uses semantic parsing algorithms to break down the query. It identifies the entities ``total sales'', ``all products'', and ``Q3''. Then, it maps these to SQL-like operations such as aggregations (e.g., SUM for calculating the total sales) and filtering operations (to select data related to Q3). Operations like SQL joins can also be synthesized when the query requires combining data from multiple tables. For instance, if the data is stored in a product table and a sales table, and the query is to find the sales of specific products, the SLM can generate a join operation to link the two tables based on a common key, such as product ID.
    \item \textbf{Enabling Complex Multi-Entity QA through Hybrid Pipelines}. The combination of these two capabilities, Relational Table Generation and Semantic Operator Synthesis, empowers the system to handle complex Multi-Entity Question Answering (Multi-Entity QA) through hybrid pipelines. Starting with unstructured data, the SLM first converts it into structured tables through Relational Table Generation. These generated tables then serve as the input for TableQA engines. For example, in a large-scale e-commerce data lake with unstructured customer reviews, product descriptions, and sales records, the SLM can transform relevant unstructured data into tables. When a complex query like ``Compare the average customer satisfaction ratings of products from different manufacturers that had a sales increase of more than 15\% in the last quarter'' is posed, the SLM-generated tables are used by the TableQA engine. The engine can then utilize the semantic operators synthesized by the SLM to perform operations like filtering the sales data for the last quarter and products with a sales increase over 15\%, and then joining the relevant tables to calculate and compare the average customer satisfaction ratings. This end-to-end process showcases how the dual-task capabilities of the SLM enable the handling of complex queries across diverse data sources.
    
\ei



\subsection{Semantic Entropy for Uncertainty Quantification}
Semantic entropy, a concept rooted in information theory, is integrated into the framework to quantify the semantic consistency of a Small Language Model’s (SLM) responses in question-answering tasks. Unlike traditional accuracy metrics, which rely on predefined ground truths, semantic entropy addresses a core challenge in natural language processing (NLP): evaluating answer quality for open-ended questions where unambiguous ``correct'' answers may not exist. By analyzing the variability in meaning across multiple generated responses (e.g., clustering answers by semantic similarity), this metric captures the model’s uncertainty and reliability. For instance, in subjective or context-dependent scenarios (e.g., legal advice or creative writing), semantic entropy reveals whether the SLM produces coherent, stable answers or diverges into conflicting interpretations—a critical measure of trustworthiness in real-world applications.

\subsubsection{Measuring Uncertainty in SLM-Generated Answers}

Semantic entropy quantifies the reliability of answers generated by a Small Language Model (SLM) by measuring consistency in meaning across multiple responses to the same input. For instance, in a medical context, if an SLM consistently answers ``What are common influenza symptoms?'' with responses like ``Fever, cough, fatigue'' and ``Symptoms include sore throat and body aches'', semantic analysis (e.g., clustering via embeddings like BERT) groups these into a single semantic cluster. Low entropy arises because all answers align with the same core meaning (influenza symptoms), indicating high reliability. This consistency reflects the model’s confidence and reduces ambiguity, making it trustworthy for critical domains like healthcare. Semantic entropy thus evaluates how reproducibly the SLM conveys factual or domain-specific knowledge, rather than judging the specificity of a single answer.

Conversely, high semantic entropy reveals uncertainty or inconsistency in the SLM’s outputs, often due to conflicting training data or ambiguous queries. For example, when asked, ``Can I be sued for sharing a photo on social media?" an SLM might generate divergent responses like ``Yes, if copyrighted'', ``No, unless consent is violated'', or ``It depends on jurisdiction''. These answers form multiple semantic clusters (e.g., "yes," "no," "conditional"), leading to high entropy. This signals unreliability, as the model fails to converge on a coherent answer, exposing gaps in its training or the query’s inherent complexity. High entropy prompts systems to flag such outputs for human review or model retraining, ensuring users receive actionable insights in domains like law, where ambiguity carries significant risk. Crucially, semantic entropy focuses on variability across responses, not the vagueness of a single reply, making it a robust metric for evaluating model confidence and contextual understanding.
\section{Conclusion and Future Work}
Our proposed SLM-driven system for unified semantic queries across heterogeneous databases simplifies data integration and demonstrates significant performance advantages in complex query scenarios. By integrating multiple advanced techniques, we have opened up a new technical path for the development of database query systems.
For future work, we plan to further optimize the retrieval mechanism to handle even larger and more diverse datasets. Additionally, we aim to explore the integration of more advanced language model architectures into the system to further enhance its semantic understanding and query answering capabilities. We also intend to expand the application domains of the system, such as applying it to real-time data analytics and knowledge database construction.
\bibliographystyle{IEEEtran}
\bibliography{refs/custom}

\end{document}